\documentclass[aps,pra,twocolumn,showpacs,showkeys,amsmath,amssymb,preprintnumbers,superscriptaddress,10pt,longbibliography]{revtex4-1}

\usepackage{amsmath}
\usepackage{graphicx}
\usepackage{ulem}
\usepackage{xurl}
\usepackage{siunitx}
\usepackage[bookmarks=false]{hyperref}
\usepackage{color}
\usepackage[capitalise]{cleveref}
\crefname{section}{Sec.}{Secs.}
\Crefname{section}{Section}{Sections}
\usepackage{makecell}

\usepackage{array}
\newcolumntype{L}[1]{>{\raggedright\let\newline\\\arraybackslash\hspace{0pt}}p{#1}}
\newcolumntype{C}[1]{>{\centering\let\newline\\\arraybackslash\hspace{0pt}}p{#1}}
\newcolumntype{R}[1]{>{\raggedleft\let\newline\\\arraybackslash\hspace{0pt}}p{#1}}

\definecolor{darkred}{RGB}{140,5,0}

\definecolor{purple}{rgb}{0.5,0.1,1}

\definecolor{pink}{RGB}{255,0,255}

\begin{document}

\title{\Large CubeSat single-photon detector module for investigating in-orbit laser annealing to heal radiation damage}

\author{Nigar~Sultana}
\email{nigareipu@gmail.com}
\affiliation{Institute for Quantum Computing, University of Waterloo, Waterloo, ON, N2L~3G1 Canada}
\affiliation{\mbox{Department of Electrical and Computer Engineering, University of Waterloo, Waterloo, ON, N2L~3G1 Canada}}

\author{Joanna~Krynski}
\affiliation{Institute for Quantum Computing, University of Waterloo, Waterloo, ON, N2L~3G1 Canada}
\affiliation{Department of Physics and Astronomy, University of Waterloo, Waterloo, ON, N2L~3G1 Canada}

\author{Jin~Gyu~Lim}
\affiliation{Institute for Quantum Computing, University of Waterloo, Waterloo, ON, N2L~3G1 Canada}
\affiliation{Department of Physics and Astronomy, University of Waterloo, Waterloo, ON, N2L~3G1 Canada}

\author{John~Floyd}
\affiliation{University of Illinois Urbana-Champaign, Champaign IL, United States}

\author{Michael~Lembeck}
\affiliation{University of Illinois Urbana-Champaign, Champaign IL, United States}

\author{Vadim~Makarov}
\affiliation{Vigo Quantum Communication Center, University of Vigo, Vigo, Spain}
\affiliation{Institute for Quantum Computing, University of Waterloo, Waterloo, ON, N2L~3G1 Canada}
\affiliation{Department of Physics and Astronomy, University of Waterloo, Waterloo, ON, N2L~3G1 Canada}

\author{Paul~Kwiat}
\affiliation{University of Illinois Urbana-Champaign, Champaign IL, United States} 

\author{Thomas~Jennewein}
\affiliation {Institute for Quantum Computing, University of Waterloo, Waterloo, ON, N2L~3G1 Canada}
\affiliation{Department of Physics and Astronomy, University of Waterloo, Waterloo, ON, N2L~3G1 Canada}
\affiliation{\mbox{Quantum Information Science Program, Canadian Institute for Advanced Research, Toronto, ON, M5G~1Z8 Canada}}

\begin{abstract}
Single-photon avalanche photodiodes (SPADs) based on silicon are widely considered for quantum satellite communications but suffer from an increasing dark count rate (DCR) due to displacement damage in their active areas induced by proton radiation. When the DCR of SPADs exceeds a certain threshold, they become unusable for quantum communication protocols. Previous laboratory experiments have demonstrated that laser annealing of SPADs' active area with about 1~W optical power can significantly reduce radiation-induced DCR of synthetically irradiated SPADs. To assess the feasibility of in-orbit laser annealing on constantly irradiated SPADs in low-Earth orbit, we developed a module with a CubeSat form factor capable of both laser and thermal annealing of four silicon SPADs. Here we report the design and ground testing of this module, investigating laser annealing in a simulated space environment. Our results pave the way for an in-orbit trial that may prove this technology useful for future satellite missions with quantum receivers on board.
\end{abstract}


\maketitle

\section{Introduction}

Single-photon detectors (SPDs) are one of the key elements in satellite-based quantum communication networks \cite{liao2017satellite, yin2017satellite, ren2017ground, yin2017satellite, liao2018satellite, takenaka2017satellite, gunthner2017quantum, yang2019spaceborne}. These networks utilize SPDs either to detect received single photons in quantum receivers or to evaluate the performance of quantum sources in orbit. Quantum receiver satellites require SPDs with low dark count rate (DCR), low timing jitter, low afterpulsing probability, and high quantum detection efficiency at wavelengths suitable for optimal atmospheric transmission, such as at \SI{800}{nm} \cite{bourgoin2013comprehensive, bourgoin2015experimental}. Commercial silicon-based single-photon avalanche photodiodes (SPADs) meet most of these requirements, with sufficiently high detection efficiency (above 60\%) over a broad range of wavelengths ($400$--$1000$\,nm). Si-SPADs have been successfully deployed in various space missions due to their long life, compact packaging, simple driving electronics, and modest cooling needs \cite{sun2004space, sun2006orbit, sun2011performance, schutz2005overview}.

However, Si-SPADs are susceptible to damage from space radiation, particularly proton radiation with energies ranging from a few MeV to several hundred MeV \cite{sun2004space}. The radiation-induced displacement damage leads to crystalline defects within the detector active regions, resulting in a constant increase in DCR, for example at a rate of approximately 30 counts per second per day, as observed at the Ice, Cloud, and land Elevation Satellite (ICESat) \cite{sun2004space}. High DCR will degrade the implementation of quantum communication protocols, such as quantum key distribution (QKD) \cite{bennett1984quantum,ekert1991quantum} and quantum teleportation \cite{bennett1993teleporting}, eventually making them impossible and thereby limiting mission lifetimes. To address this issue, previous ground studies have demonstrated the effectiveness of mitigating radiation-induced damage through thermal \cite{sun1997measurement,tan2013silicon,sun2001proton,marisaldi2011single,prochazka2007single,moscatelli2013radiation,anisimova2017mitigating} or laser annealing \cite{lim2017laser, bugge2014laser} procedures. Particularly, laser annealing using a controlled amount of  laser light  has been found to be  more effective in reducing DCR by healing crystalline defects through focused heating of detector active areas \cite{lim2017laser}. However, the in-orbit effectiveness of laser annealing on continuously irradiated Si-SPADs in low Earth orbit (LEO) remains unknown. Therefore, our study aimed to develop technology for evaluating the efficacy of in-orbit laser annealing on continuously irradiated Si-SPADs, with the goal of extending mission lifetimes.

To facilitate this study, a detector (SPAD) module has been developed for the CubeSat mission CAPSat (Cool Annealing Payload Satellite) \cite{UIUC}, and a similar design has been employed for the new satellite mission Space Entanglement and Annealing QUantum Experiment (SEAQUE) \cite{SEAQUE}. This paper presents a detailed design of our SPAD module, which enables thermal and laser annealing and detector characterization in orbit. The module employed a novel approach to fiber-coupling the annealing lasers to the SPADs, ensuring full illumination of the detector active areas. We demonstrated the module's operation in a simulated space environment inside a thermal vacuum chamber (TVAC), representing a significant step toward establishing in-orbit laser annealing of SPADs. This paper also presents performance test results for the detectors from the flight SPAD module and for laser-annealed irradiated detectors from an equivalent flight-compatible engineering detector module.

\section{Module design}

\begin{figure}
\centering
	\includegraphics{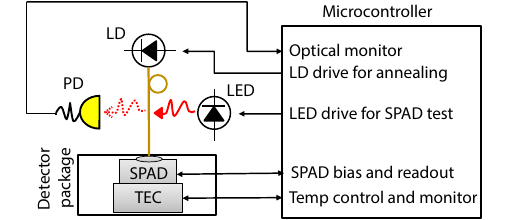}
	\caption{Concept of the in-orbit laser annealing. A high-power beam from a laser diode (LD) is transmitted through an optical fiber and focused onto the active area of a SPAD. A light emitting diode (LED) acts as a reference light source for testing the detectors' photon detection efficiency. A photodiode (PD) is used to monitor the optical power from the LED; a thermoelectric cooler (TEC) integrated with the SPAD is employed for temperature regulation during operation, and a microcontroller controls the operation of all the devices.}
	\label{fig:Laser_annealing}
\end{figure}

A functional representation of an in-orbit laser annealing system is illustrated in  \Cref{fig:Laser_annealing}. The system employs a high-power annealing laser diode (LD) to illuminate SPAD active areas through an optical fiber. For verification of photon-detection efficiency, an adjacent light-emitting diode (LED) is used to emit light, which leaks into the jacketed fiber, primarily targeting the core of the fiber. The power output of the LED is monitored and measured using an additional photodiode (PD). The operation of all the components is intended to be controlled by a microcontroller (MCU).  The MCU is also expected to collect telemetry data during in-orbit operations. This telemetry includes various metrics such as detector dark counts, temperature readings, TEC current, PD photocurrent, and other relevant measurements. Data collection can occur under different cooling temperatures, before and after each annealing cycle, and following a number of orbital revolutions, which can vary depending on the dark count (DCR) accumulation rate. The collected telemetry is subsequently transferred to the onboard computer for storage, until it can be transmitted to a ground station. Comprehensive experimental laser annealing protocols in terms of annealing laser power,  annealing duration are analyzed in \cite{krynski2023protocols}.

\begin{figure}
	\centering
	\includegraphics[width=0.70\linewidth, height=0.27\textheight]{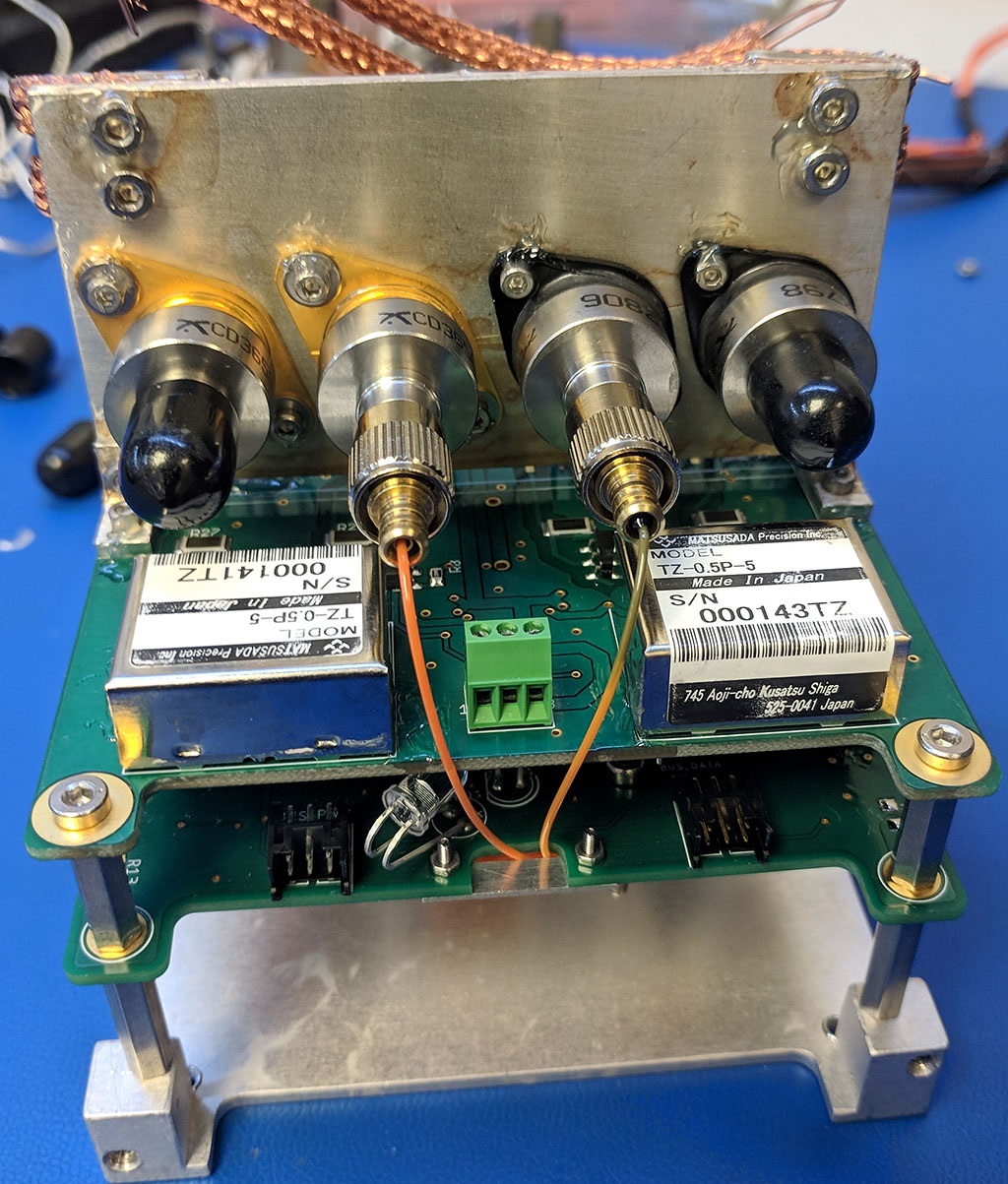}
	\caption{SPAD module integrated with the CAPSat control module.}
	\label{fig:DM2_ControlBoard}
\end{figure}

The laser annealing system built for the CAPSat is divided into two modules. The detectors and their associated circuitry required for performing in-orbit laser annealing and thermal annealing are housed in the SPAD module, while the other essential components for the annealing operation—such as the fiber-pigtailed LDs, LEDs, PDs, and the MCU—are located in a separate module known as the control module. \Cref{fig:DM2_ControlBoard} illustrates the two integrated modules of the CAPSat annealing system.  This paper focuses solely on the SPAD module and provides its  detailed design and performance results.

\begin{figure}
	\centering
	\includegraphics{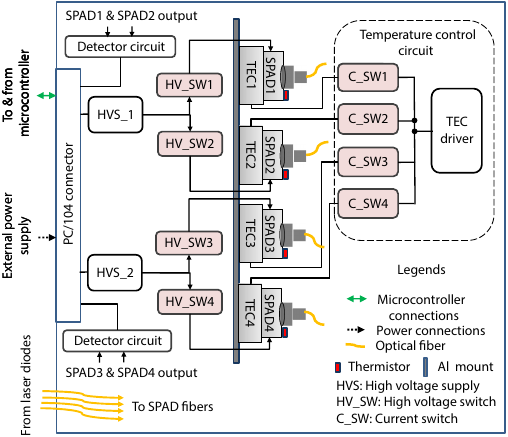}
	\caption{Schematic overview of the CAPSat SPAD module, highlighting the main components and functional elements.}
	\label{fig:Module_schematic}
\end{figure}

The SPAD module consists of two Excelitas SLiK and two Excelitas C30902SH detectors, each with active area diameters of \SI{180}{\micro\meter} and \SI{500}{\micro\meter}, respectively. This module also includes detector biasing, quenching, pulse shaping and temperature control circuitry, a schematic overview is illustrated in \cref{fig:Module_schematic}. Each detector package, along with a SPAD, also includes a built-in two-stage thermoelectric cooler (TEC) and a thermistor used for detector temperature control and monitoring. The detectors are customized to have an attached fiber connector through which  annealing lasers can be coupled via a multimode optical fiber, ensuring effective illumination of the entire active area. In CAPSat, only one SLiK and one  C30902SH detectors are fiber-coupled with a core diameter of \SI{105}{\micro\meter} with the two laser diodes of the control module for laser annealing, while the other two detectors  from each variant remain as control devices that do not undergo laser annealing and are intended to measure the overall damage caused by space radiation.

 The detectors are mounted on an aluminum plate, which serves as both a physical support and a heat sink. This plate is vertically attached  to the module printed circuit board (PCB), which also holds two high-voltage supplies (HVS) to provide bias voltages, and two detector circuits consisting of quenching and pulse-shaping electronics for processing avalanche signals. Each circuit is shared by two SPADs. Only one SPAD is selectively activated during operation via a high-voltage switch (HV\_SW, Panasonic Electric Works AQV216SX). Additionally, the PCB incorporates a TEC driver and four current switches (C\_SW, Omron Electronics G3VM-21HR) to selectively control the current flow through one of the four TECs to maintain desired detector temperatures. This approach provides the flexibility to operate a single detector and a corresponding TEC at any given time, effectively minimizing the power consumption of the SPAD module. The power and control signals for the SPAD module are supplied from the control module via a PC/104 connector. At the development and testing phase of this module, a Cypress controller (CY8CKIT-030 PSoC) was used to supply the control signals.

\begin{figure}
	\includegraphics{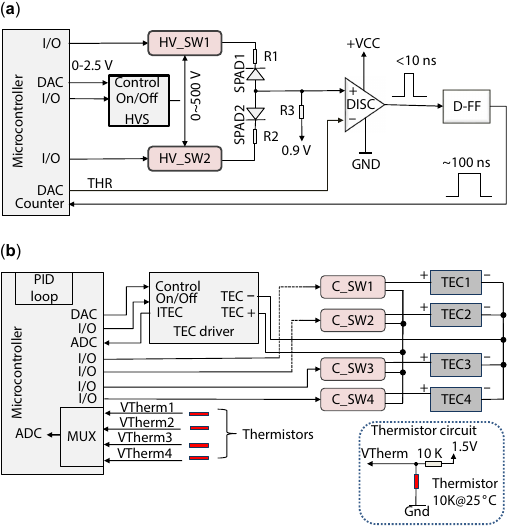}
	\caption{Simplified circuit diagrams of detector electronics. (a)~Detector biasing and quenching electronics. Two detectors share the same high-voltage supply (HVS) and passive quenching and pulse-shaping circuits. A microcontroller (MCU) controls HVS output, detector activation one at a time, discriminator threshold voltage, and records detector counts. (b)~Temperature control circuits. A driver controls current flow and its direction through the TECs using feedback from a PID loop running on the MCU based on thermistor readings. The microcontroller activates one TEC at a time, reads thermistor values, and sends control signals to the TEC driver.}
	\label{fig:Circuits}
\end{figure}

\begin{figure*}
\centering
	\includegraphics{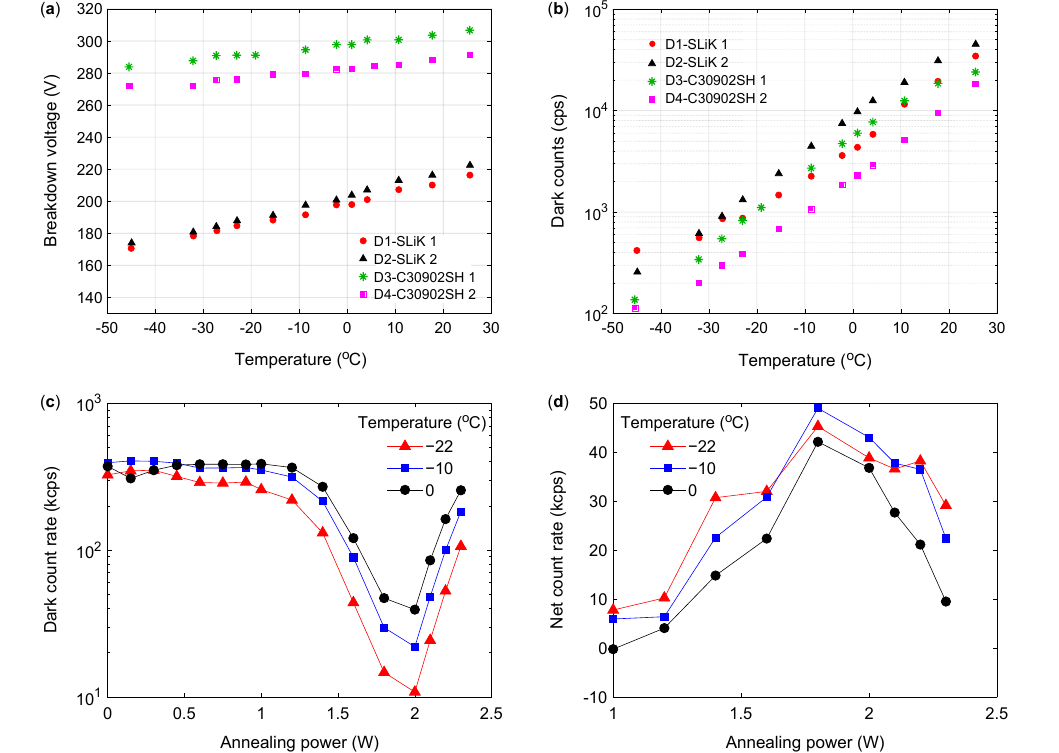}
	\caption{Results of performance tests. (a)~Breakdown voltage and (b)~dark count rate of SPADs in the flight annealing payload across the temperature range. (c,d)~Data from TVAC testing of the engineering module, showing (c)~measured dark count rate and (d)~net count rate of an irradiated SPAD  after its annealing with increasing powers. Plot (c) and \SI{-22}{\celsius} data in (d) are reprinted from \cite{krynski2023protocols}.}
	\label{fig:detector_parameter}
\end{figure*}

The detectors of the SPAD module operate in Geiger mode \cite{haitz1965mechanisms} above the breakdown voltage $\text{V}_b$, generating an avalanche current in response to a single photon. This current is halted by a passive quenching circuit that includes a quenching resistor R1 or R2 of \SI{402}{\kilo\ohm} \cite{cova1996avalanche}, as shown in the detector circuit in \cref{fig:Circuits}(a). A digital-to-analog converter (DAC) internal to the MCU provides a control voltage ranging between $0$--$2.5$\,V for the HVS that outputs biasing voltages ranging between $0$--$500$\,V. When a SPAD is biased, each detection event creates a voltage drop across R3 (\SI{100}{\ohm}), the magnitude of which is compared with a MCU-DAC-provided predefined threshold voltage (THR) using a high-speed radiation-tolerant comparator (DISC, STMicroelectronics RHR801). As the module supports only single-polarity supplies instead of the comparator's requirement of dual-polarity power supply, the inputs must be within a range of $0.5$--$2.1$\,V with the use of VCC\,=\,$+3.3$\,V supply. Therefore, a reference of $900$\,mV is used instead of ground, using a voltage reference IC (Intersil ISL21070). When the voltage across R3 is higher than the predefined threshold, the comparator generates $3$\,V output pulses of $5$--$10$\,ns width. Their width is then increased to approximately $100$\,ns using a D flip-flop (D-FF; Texas Instruments 74AC11074pwr) so that they can be read out by the relatively slow MCU. A counter configured in the MCU then records the number of detection events within a measurement time period.

In order to minimize dark counts, and maintain a consistent detection efficiency at a fixed bias voltage, the SPADs are typically cooled to a fixed temperature, which is controlled using their built-in TECs. The latter can also be utilized for thermal annealing by reversing the direction of the TEC current [\cref{fig:Circuits}(b)]. To control the current flow and direction through the TECs, a compact TEC driver (Maxim Integrated MAX1968) is employed, which receives feedback from a proportional integral derivative (PID) loop programmed in the MCU. The PID algorithm reads the instantaneous detector temperatures as a voltage drop across the built-in thermistors (VTherm) via an MCU analog-to-digital converter (ADC). Based on the difference between the instantaneous and desired temperatures, the PID algorithm sends appropriate control voltages to the TEC driver to adjust its output current, which flows through the active TEC. The current flow is monitored by observing the ITEC readout of the TEC driver using the MCU ADC.

\section{Performance tests}

\textbf{Pre-radiation SPAD performance test:} 
This test was performed at atmospheric pressure on the flight model of the CAPSat SPAD module that included all-new detector samples. The breakdown voltages and DCR of the four SPADs were measured, see \cref{fig:detector_parameter}(a,b). The SLiK devices (D1 and D2) have lower breakdown voltages in comparison to the C30902SH samples (D3 and D4). The breakdown voltage for SLiK (C30902SH) devices varied by approximately \SI{0.6}{\volt\per\celsius} (\SI{0.3}{\volt\per\celsius}). As temperature increased, the DCR of the detectors increased exponentially up to tens of kilocounts per second when biased at \SI{20}{\volt} excess bias [\cref{fig:detector_parameter}(b)]. 

\medskip 

\begin{figure}
	\centering
	\includegraphics[width=.5\textwidth]{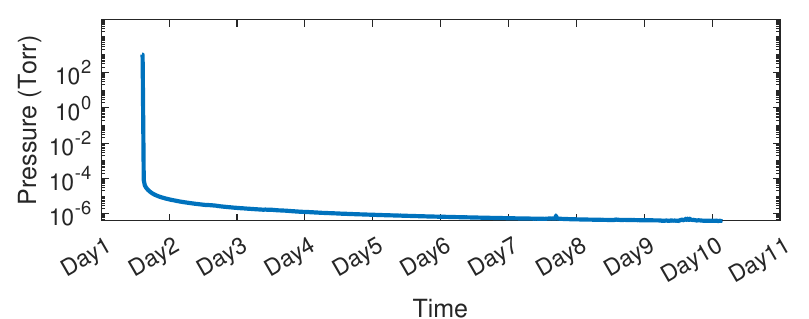}	
	\caption{Observed pressure inside TVAC chamber over 10 days of the SPAD module operation.}
	\label{fig:pressure}
\end{figure}

\textbf{Post-radiation SPAD performance test:} This test was conducted inside a thermal vacuum chamber at a pressure of $10^{-6}$\,Torr on an engineering model (EM) of the SPAD module. This EM module is equivalent to the flight module but utilizes four pre-irradiated SLiK detectors, each subjected to a ten-year equivalent low Earth orbit proton dose of $2\times10^{10}$ protons/cm$^2$, as reported in \cite{dsouza2021repeated}.  This test was performed to verify the expected behaviour of this module in a space environment. All four detectors underwent laser annealing at vacuum pressure, utilizing optical power from a laser diode located outside the TVAC chamber. Detailed test procedures and protocols are described in \cite{krynski2023protocols}. This test demonstrated that high annealing power applied for a short duration can effectively reduce the dark counts. For instance, \cref{fig:detector_parameter}(c,d) illustrates that optical power ranging from 1\,W to 2\,W for \SI{180}{\second} improved both the dark count rates and net count rates (the difference between the count rates with and without light). For more annealing results, see \cite{krynski2023protocols}.

\medskip 

\begin{table}[t]
	\caption{Key features of the SPAD module in TVAC testing}
	\label{tab:DM1_SPADS}
	\footnotetext[1]{Measured at room temperature, with only one detector and one TEC running at a time.}
	\footnotetext[2]{During \SI{\sim 1}{h} of operation under vacuum.}
	\begin{tabular}[t]{L{37.7mm}L{46mm}}
	\hline\hline      
Size & $90 \times 90 \times 52~\text{mm}^3$ \\
Mass & \SI{250}{g} \\
Component grade & Mostly automotive or industrial \\ 
Power consumption & \makecell[lt]{\SI{0.02}{W} in quiescent mode\\ \SI{0.93}{W} with SPAD at \SI{4}{\celsius}\footnotemark[1]\\ \SI{1.28}{W} with SPAD at \SI{-23}{\celsius}\footnotemark[1]} \\
Maximum temperature increase\footnotemark[2] & \makecell[lt]{\SI{4.56}{\celsius} (detector aluminum mount)\\ \SI{4.12}{\celsius} (high-voltage supply)\\ \SI{4.69}{\celsius} (bottom of PCB)} \\
	\hline\hline      
	\end{tabular}
\end{table}

\textbf{Engineering performance of the module:} Key features of this module are provided in \cref{tab:DM1_SPADS}.  Its total power consumption is under \SI{1.3}{W}, well below the CAPSat requirement of \SI{2}{W}.  During TVAC testing, no noticeable outgassing events were observed, as shown in (\cref{fig:pressure}). Continuous temperature monitoring was conducted at three locations on the SPAD module: the aluminum mount, one of the high voltage supplies, and the bottom of the PCB during repeated annealing and subsequent operation of the module. The highest recorded temperature was  \SI{5}{\celsius} above ambient at the aluminum mount, with no significant temperature fluctuations observed during extended SPAD operation inside vacuum, demonstrating its capability to function effectively in a vacuum environment.

\section{Conclusion}

We presented a detailed design and implementation of the detector module, which was specifically engineered to function efficiently in the low-power and low-mass environment typical of CubeSats. We have also explained the concept of in-orbit operation of the detector module. We have completed its ground tests in preparation to implementing laser annealing on SPADs in space. The design of our detector module has been adapted for the upcoming Space Entanglement and Annealing QUantum Experiment (SEAQUE) \cite{SEAQUE}, which will demonstrate the in-orbit annealing along with an entangled photon source, showcasing the potential advancement of space-based SPAD technology.
\medskip 

\section*{Acknowledgements}
We thank Excelitas Technologies for discussions, customising their device package, and providing selected SPAD samples. This work is funded by the Canada Foundation for Innovation (CFI), the Canadian Space Agency (CSA), the Ontario Research Fund (ORF), the Natural Sciences and Engineering Research Council of Canada (NSERC), and Industry Canada.

\bibliographystyle{unsrt}

\end{document}